\documentstyle[prl,twocolumn,floats,epsf,aps]{revtex}
\begin{document}
\draft

\twocolumn[\hsize\textwidth\columnwidth\hsize\csname @twocolumnfalse\endcsname

\title{Spin-Peierls transition in the Heisenberg chain
with finite-frequency phonons}

\author{Anders W. Sandvik and David K. Campbell}
\address{Department of Physics, University of Illinois at Urbana-Champaign,
1110 West Green Street, Urbana, Illinois 61801 \cite{permanent} \\
and Center for Nonlinear Studies, Los Alamos National Laboratory,
Los Alamos, New Mexico 87545}

\date{\today}

\maketitle

\begin{abstract}
We study the spin-Peierls transition in a Heisenberg spin chain coupled 
to optical bond-phonons. Quantum Monte Carlo results for systems with up to 
$N=256$ spins show unambiguously that the transition occurs only when the 
spin-phonon coupling $\alpha$ exceeds a critical value $\alpha_c$. Using 
sum rules, we show that the phonon spectral function has divergent (for
$N \to \infty$) weight extending to zero frequency for $\alpha < \alpha_c$.
The equal-time phonon-phonon correlations decay with distance $r$ as $1/r$. 
This behavior is characteristic for all $0 < \alpha < \alpha_c$ and the 
$q=\pi$ phonon does not soften (to zero frequency) at the transition.
\end{abstract}


\vskip2mm]

The $S=1/2$ Heisenberg spin chain is unstable towards dimerization 
(the spin-Peierls transition) when coupled to an elastic lattice 
\cite{crossfisher}. For phonons in the adiabatic limit, 
this transition has been predicted to occur for arbitrarily weak spin-lattice
coupling \cite{crossfisher}. On the other hand, recent work for optical
phonons in the anti-adiabatic (high-frequency) limit suggests a transition 
only above a critical coupling \cite{uhrig}. Furthermore, the mechanism of 
the transition in this limit was suggested to be qualitatively different, 
with no softening of the $q=\pi$ phonon \cite{uhrig}. A way to reconcile 
the results in the adiabatic and anti-adiabatic limits has been proposed 
within an improved mean-field (RPA) theory \cite{gros}, with the result that
a complete phonon softening occurs only for bare phonon frequencies 
$\omega_0$ less than a critical value. For higher $\omega_0$, a central
peak appears in the phonon spectral function and the phonon branch remains 
gapped. Considering the manifestly uncontrolled nature of mean-field 
calculations in one dimension, non-perturbative results in the regime of 
phonon frequencies comparable to the magnetic exchange energy $J$ are 
required to test this novel scenario.

In this Letter, we address the issues of a critical spin-phonon coupling and
the mechanism of the zero temperature dimerization transition in the strictly
one-dimensional case, using quantum Monte Carlo (QMC) simulations 
to obtain numerically exact results for relatively large systems. 
The model we study is defined by the Hamiltonian
\begin{equation}
H = J\sum\limits_{i=1}^N (1+\alpha x_i){\bf S}_i \cdot {\bf S}_{i+1}
+ \omega_0 \sum\limits_{i=1}^N n_i,
\label{hamiltonian}
\end{equation}
where 
\begin{equation}
x_i = (a^+_i + a_i)/\sqrt{2}
\label{xdef}
\end{equation}
is the phonon coordinate, and $n_i=a^+_ia_i$ is the phonon occupation number 
at bond $i$. We use a recently developed QMC method based on sampling the 
perturbation expansion in the interaction representation \cite{irsse}. For 
a finite lattice at finite inverse temperature $\beta$, the expansion 
converges for any decomposition of $H = H_0 + V$ into diagonal 
($H_0$) and perturbing ($V$) terms and can be
used \cite{prokofev} as a basis for a ``worldline'' Monte Carlo algorithm 
in continuous imaginary time (i.e., without invoking the Trotter decomposition
\cite{worldline}). In a slight modification of the scheme introduced in 
Ref.~\onlinecite{irsse}, we here include only the bare phonons in the 
diagonal term; $H_0=\omega_0\sum_i n_i$. The updating of the spin degrees of 
freedom can then be carried out using a new and highly efficient 
``operator-loop'' algorithm \cite{sandvik}, which in particular allows for 
sampling of all winding number sectors and hence direct evaluation of the spin 
stiffness \cite{pollock}. 

In this work we consider only an energy $\omega_0=J/4$ for the bare phonons 
and study the behavior for values of the spin-phonon coupling in the range 
$0 \le \alpha/J \le 0.5$. We have studied systems with $N$ up to 
$256$ at inverse temperatures $\beta=J/T$ sufficiently high to give 
ground state results. Typically, for the system sizes we have considered, 
$\beta$ as high as $\approx 2N$ is required to achieve convergence to 
the $T=0$ limit of all the quantities of interest. We have used at 
least $\beta = 4N$ for all calculations presented here. 

Note that in the model, Eq.~(\ref{hamiltonian}), for $\alpha > 0$ 
there is an energy gain associated with an average uniform phonon displacement 
$\langle x \rangle  = (1/N)\sum_i \langle x_i \rangle > 0$, which leads to an 
increased average effective spin-spin coupling $J_{\rm eff} = J + \alpha 
\langle x \rangle > J$. For $\omega_0/J = 0.25$ at $T=0$, we find 
$J_{\rm eff}/J =  1.018, 1.071, 1.158, 1.278$, and $1.430$ for $\alpha = 
0.1,0.2,0.3, 0.4$, and $0.5$, respectively. We will in some cases measure 
energies in units of $J_{\rm eff}$ instead of the bare exchange $J$.

The most direct signal of the dimerization that can be measured in our
simulations is the approach of the staggered phonon-phonon correlation
function $(-1)^r \langle x_{i}x_{i+r}\rangle$ to a non-zero value at long 
distances $r$. Our results for this quantity indicate a critical coupling
$0.1 <\alpha_{\rm c}/J < 0.35$ for $\omega_0/J=0.25$ \cite{icsm}. In order 
to improve on the accuracy of this rough estimate, and to circumvent
potential problems with detecting a very small dimerization (as would be 
the case for very weak coupling if the mean-field result $\alpha_{\rm c} 
= 0$ would be correct), we have also considered several other quantities. It
is particularly useful to study the effects of the dynamic phonons in the spin 
sector (in general, our simulation results for spin quantities have smaller 
statistical fluctuations than the phonon correlations). Thus we
discuss here results for the spin stiffness and the staggered spin 
susceptibility.

\begin{figure}
\centering
\epsfxsize=8cm
\leavevmode
\epsffile{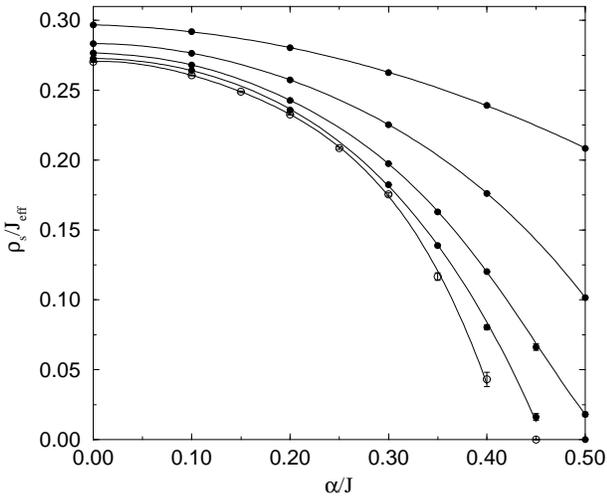}
\vskip1mm
\caption{Spin stiffness vs the spin-phonon coupling for system sizes
$N=8,16,32,64$ and $128$ ($\rho_s$ decreases with increasing $N$). Where 
not shown, statistical errors are smaller than the symbols. The curves 
are high-order polynomial fits to the data points.}
\label{fig1}
\end{figure}

If the dimerization transition occurs at some critical coupling 
$\alpha_{\rm c} > 0$, it is expected to be of the Kosterliz-Thouless (KT)
type \cite{kt}. The spin stiffness $\rho_{\rm s}$, i.e., the ground state 
energy curvature with respect to a uniform twist $\phi$ in the
spin-spin interaction \cite{shastry},
\begin{equation}
\rho_{\rm s} = {1\over N} {\partial^2 E_0 (\phi) \over \partial \phi^2 },
\end{equation}
is then expected to exhibit a discontinuous jump from a finite value
for $\alpha \le \alpha _{\rm c}$ to zero for $\alpha > \alpha _{\rm c}$
(reflecting the opening of a spin gap). For a finite system the jump will
be smoothed. In Figure \ref{fig1} we show results for the stiffness versus
$\alpha/J$ for several system sizes. The behavior expected for a KT 
transition is seen clearly --- $\rho_s$ rapidly approaches zero for
$\alpha/J \agt 0.4$ but appears to converge to a finite value for
$\alpha/J \alt 0.2$, indicating a critical coupling between these
values, in agreement with our previous results for the dimerization.
It is, however, not easy to extract an accurate value for $\alpha_{\rm c}$ 
using these results. The scaling behavior is complicated by logarithmic 
corrections, which we expect to be present for all $\alpha \le \alpha_c$
as in the case of the Heisenberg chain [i.e., $\alpha = 0$ in 
Eq.~(\ref{hamiltonian})]. This is in contrast to the finite temperature 
KT transition in the two-dimensional XY model, where $\rho_s$ approaches 
its asymptotic value algebraically for $T < T_{\rm c}$ and logarithmically 
only exactly at $T_{\rm c}$ \cite{minnhagen,note}. An indication of the 
difficulties associated with the log corrections in the spin-phonon chain 
can be seen in our stiffness 
data for $\alpha =0$, for which the exact infinite-size value is known to
be (in our units) \cite{shastry} $\rho_s=1/4$; about $8\%$ lower than
what we find for $N=128$.

\begin{figure}
\centering
\epsfxsize=8cm
\leavevmode
\epsffile{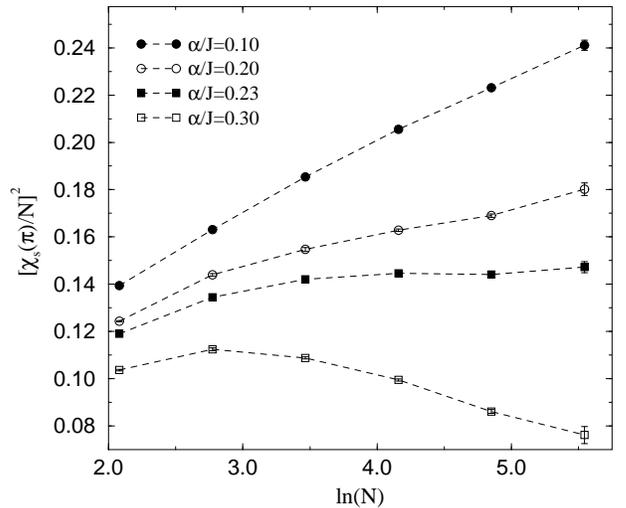}
\vskip1mm
\caption{Size dependence of the staggered spin susceptibility for different
values of the spin-phonon coupling. A linear behavior of 
$[\chi_s(\pi)/N]^2$ vs $\ln (N)$ is expected in the gapless phase, 
with slope zero at the critical point. The decrease with increasing 
$N$ for $\alpha/J=0.3$ indicates the presence of a spin gap.}
\label{fig2}
\end{figure}

Although the log corrections complicate the extraction of $\alpha_{\rm c}$
from the stiffness data, their presence in other quantities can in
fact be useful in numerical calculations. The asymptotic behavior of 
the spin-spin correlation function of the Heisenberg chain is known 
from bosonization and conformal field theory \cite{luther,logpapers};
\begin{equation}
\langle S^z_i S^z_{i+r} \rangle \sim {(-1)^r\over r} {\rm ln}^{1/2} 
(r/r_0).
\label{spincorr}
\end{equation}
We expect this form to apply for all $\alpha < \alpha_c$. The logarithmic 
correction should vanish at the critical point $\alpha_c$, as it is known to 
do, e.g., at the critical point of the frustrated $J_1-J_2$ chain 
\cite{eggert}. We have calculated the staggered spin susceptibility
\begin{equation}
\chi_s (\pi) = {1\over N} \sum\limits_{m,n} (-1)^{n-m} \int_0^\beta
d\tau \langle S^z_n(\tau)S^z_m(0) \rangle ,
\end{equation}
for which Eq.~(\ref{spincorr}) and conformal invariance imply the
finite-size scaling form \cite{logpapers}
\begin{equation}
\chi_s (\pi) \sim N{\rm ln}^{1/2}(N/N_0).
\label{chiscaling}
\end{equation}
In Figure \ref{fig2} we graph $(\chi_s(\pi)/N)^2$ vs $\ln (N)$ for $\alpha/J$ 
in the range $0.1 - 0.3$. For $\alpha/J=0.1$ and $0.2$ the linear behavior 
for the larger system sizes is consistent with the form (\ref{chiscaling}) 
expected in the gapless phase, whereas in the $\alpha/J = 0.3$ case there is a 
clear decrease with increasing $N$, corresponding to a finite asymptotic 
value for $\chi_s(\pi)$ and therefore the presence of a spin gap. 
For $\alpha/J=0.23$
the curve is flat within statistical errors for $N \ge 64$, implying that
$\chi_s (\pi)$ diverges linearly with $N$ without log correction. Based on 
this result (and calculations for other $\alpha/J$ close to $0.23$) we 
conclude that $\alpha_c/J = 0.23 \pm 0.01$.

\begin{figure}
\centering
\epsfxsize=8cm
\leavevmode
\epsffile{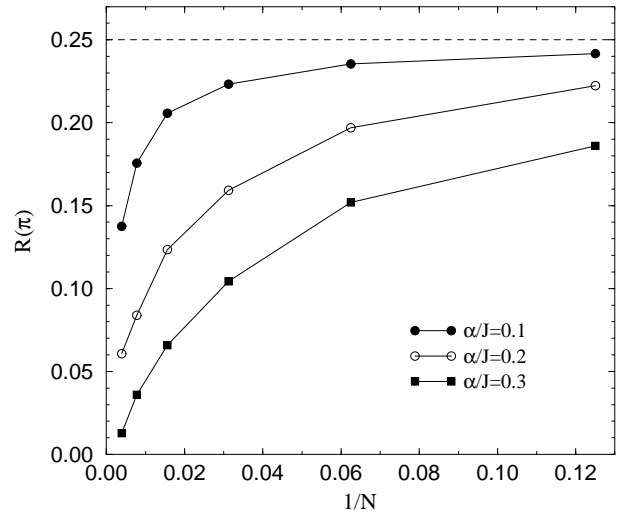}
\vskip1mm
\caption{Upper bound $R(\pi)=2S_x(\pi)/\chi_x(\pi)$ for the lowest $q=\pi$
phonon excitation energy vs the inverse system size. The dashed line
is at the non-interacting ($\alpha=0$) value $\omega_0/J$.}
\label{fig3}
\end{figure}

Having established a KT transition and the critical coupling, we now turn 
to the question of the behavior of the $q=\pi$ phonons at the transition.
We consider the phonon spectral function
\begin{equation}
A(q,\omega) = \sum\limits_{m,n} {\rm e}^{-\beta E_n}
|\langle m | x_q | n\rangle |^2 \delta (\omega - [E_m-E_n]) ,
\label{aqw}
\end{equation}
where 
\begin{equation}
x_q = {1\over \sqrt{N}}\sum_{j=1}^N {\rm exp}(-iqj) x_j.
\end{equation}
This real-frequency dynamic quantity cannot be obtained directly in our 
simulations. In order to avoid the problems associated with numerically 
continuing imaginary time data to real frequency, we here study sum rules 
that relate $A(q,\omega)$ to quantities that can be directly calculated. 
Two useful integrals that can be easily obtained from Eq.~(\ref{aqw}) are
\begin{mathletters}
\begin{eqnarray}
S_x(q) && = 
\int_0^\infty d\omega A(q,\omega) (1+{\rm e}^{-\beta\omega}), \label{sums} \\
 \chi_x(q) && = 
2\int_0^\infty d\omega A(q,\omega)\omega^{-1} \label{sumx}
(1-{\rm e}^{-\beta\omega}) ,
\end{eqnarray}
\end{mathletters}
where $S_x(q)$ and $\chi_x (q)$ are the static structure factor and
susceptibility;
\begin{mathletters}
\begin{eqnarray}
S_x(q) && = \langle x_{-q} x_q \rangle \\
\chi_x(q) && = \int_0^\beta d\tau \langle x_{-q}(\tau) x_q (0) \rangle .
\end{eqnarray}
\end{mathletters}
Using Eqs.~(\ref{sums}) and (\ref{sumx}) one can readily verify \cite{cuo}
that the ratio
\begin{equation}
R(q) = 2S_x(q)/\chi_x(q)
\end{equation}
is an upper bound for the lowest phononic excitation of momentum $q$. For
$\alpha \ge \alpha_c$ we therefore expect $R(\pi) \to 0$ as $N \to \infty$,
reflecting the presence of two degenerate ground states with momenta
$0$ and $\pi$ (linear combinations of the two possible real-space 
dimerized states). For $\alpha=0$, $R(q)=\omega_0$ for all $q$. A 
transition caused by a softening of the $q=\pi$ phonon would imply 
$R(\pi) > 0$ for $\alpha < \alpha_c$ and $R(\pi) \to 0$ as 
$\alpha \to \alpha_c$. This behavior is not seen in our results. 
Instead, $R(\pi)$ appears to approach zero as $N$ is increased even 
for $\alpha$ much smaller than the critical value, as shown in 
Figure~\ref{fig3}. Hence {\it the spectral weight extends to zero frequency 
also in the non-dimerized systems}. 

\begin{figure}
\centering
\epsfxsize=6.5cm
\leavevmode
\epsffile{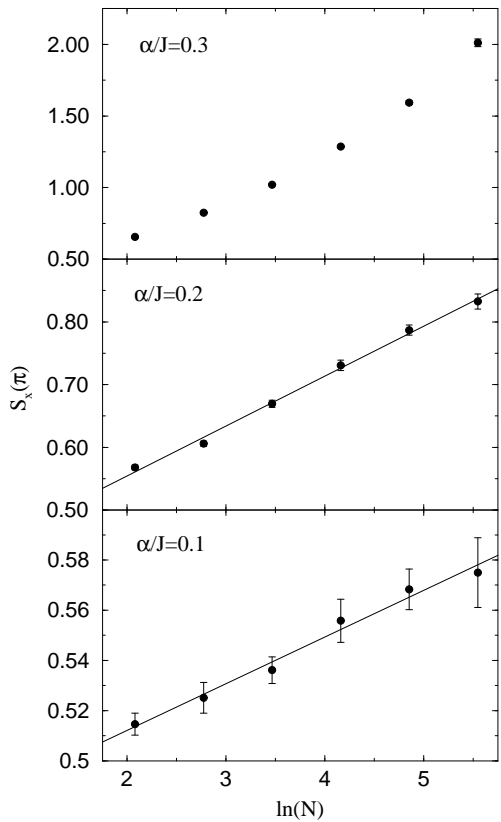}
\vskip1mm
\caption{Staggered phonon structure factor vs the logarithm of the system 
size. The linear behavior for $\alpha=0.1$ and $0.2$ corresponds to an inverse 
distance decay of the real-space phonon correlation function. Note
that with our definition [Eq.~(\protect{\ref{xdef}})] of the phonon 
coordinate, the non-interacting structure factor equals $0.5$.}
\label{fig4}
\end{figure}

We also find that the total spectral weight, given by the structure factor 
according to Eq.~(\ref{sums}), diverges with $N$. In Figure~\ref{fig4} we 
graph $S_x(\pi)$ versus the logarithm of the system size for values of
$\alpha$ both below and above the critical coupling. We find a linear
increase for $\alpha < \alpha_c$, indicating a logarithmic divergence
and therefore an inverse distance decay of the real-space phonon-phonon 
correlation function. For $\alpha = 0.3 > \alpha_c$ the divergence 
is faster than logarithmic. The expected linear in $N$ behavior cannot
be observed close to $\alpha_c$ for the system sizes we have studied,
due to large short-distance contributions to $S_x (\pi)$. For 
$\alpha \ge 0.4$ we do observe an almost linear divergence with $N$.

These results show that, in the thermodynamic limit, there is infinite
$q=\pi$ phonon spectral weight also for $\alpha < \alpha_c$. This weight 
extends to zero frequency. The rate of decay of $R(\pi)$ with increasing 
$N$, seen in Fig.~\ref{fig3}, shows that the low-frequency weight grows 
rapidly with $N$. The only plausible explanation for this is that the 
phonon spectral function has a central peak with infinite integral. We 
now elaborate on the reasons for this behavior.

In the absence of spin-phonon couplings, the low-lying excitations of the 
system are the two-spinon singlet and triplet states of the Heisenberg chain. 
Our results indicate that an arbitrarily weak coupling to the phonons induces
an infinite phonon spectral weight into these states at the staggered 
momentum $q=\pi$. The asymptotic real-space staggered phonon correlation 
function has the same $1/r$ decay as the spin-spin correlation function 
(perhaps differing by multiplicative logarithmic corrections that cannot 
be detected in our results for the phonon correlations). This is 
not completely surprising, considering that the Heisenberg chain is also  
characterized by an inverse distance decay of the dimerization correlation 
function $\langle (S_i \cdot S_{i+1})(S_{i+r}\cdot S_{i+1+r})\rangle$, 
as also recently noted by Gros and Werner \cite{gros}. The corresponding 
susceptibility is therefore divergent and this leads to the spontaneous 
dimerization for arbitrarily weak spin-phonon couplings in the adiabatic
case $\omega_0 = 0$. What we have shown here is that dynamic ($\omega_0 > 0$)
phonons destroy the long-range order for weak spin-phonon couplings but 
nevertheless the spin and phonon excitations remain coupled in a 
manifestly non-perturbative fashion.

The most plausible scenario for the mechanism of the spin-Peierls transition
is then the following: For weak coupling $\alpha$, the phonon spectral function
at $q=\pi$ has a finite-weight peak close to the bare frequency $\omega_0$, 
as well as a central peak with divergent frequency integral
[$\ln{(N)}$ divergent as a function of system size]. As $\alpha$ is 
increased the finite-frequency peak may shift slightly but remains at finite
frequency. The central peak sharpens and at $\alpha = \alpha_c$ acquires 
a $\delta$ function component, corresponding to the development of static
long-range order. For $\alpha > \alpha_c$ this $\delta$ function remains,
and a gap develops to the remainder of what was the finite-width central 
peak for $\alpha \le \alpha_c$. This gap is the excitation energy of a
lattice/magnetic soliton pair. 

On general grounds, we find it unlikely that there would be any qualitative 
changes in the nature of the $T=0$ transition as $\omega_0$ is varied. 
A recent density matrix renormalization group study of a model closely 
related to the one we have studied also finds unambiguously that the 
critical coupling $\alpha_c > 0$ for any $\omega_0$ \cite{bursill}. In
addition, our results are consistent with the calculations in the 
anti-adiabatic limit \cite{uhrig} (although the infinite low-frequency 
$q=\pi$ phonon weight for $\alpha < \alpha_c$ was not noted there), 
even though our $\omega_0=J/4$ is closer to the adiabatic regime. 

We have here discussed only the $T=0$ quantum phase transition in the
strictly one-dimensional case. The finite $T_c$ in real materials such
as CuGeO$_{\rm 3}$ \cite{hase} can be due to three-dimensional
phonons, as well as interchain magnetic couplings. The non-softening 
nature of the quantum phase transition that we have found here clearly 
supports the suggestion \cite{uhrig,gros} that the finite-$T$ transition 
may also be non-softening. In the improved RPA theory \cite{gros}, 
softening occurs below a critical value of $\omega_0$. In future work, 
we plan to extend our simulations to two- and three-dimensional systems 
and hope to address this important issue.

We would like to thank  Robert Bursill, Valeri Kotov, Ross McKenzie,
Oleg Sushkov, and Johannes Voit for useful discussions. The numerical 
simulations were carried out at the NCSA facilities at the University 
of Illisois at Urbana-Champaign. This work is supported by the National 
Science foundation under Grant No.~DMR-97-12765.


\begin{references}
\bibitem{note}
It can be noted that the isotropic Heisenberg model also corresponds to
a KT point of the transition between XY-like and Ising-like behavior
in the axially anisotropic model. Such a transition should also be present at
the isotropic point for $\alpha > 0$. Hence, in an extended anisotropic
spin-phonon model there is a line of KT points. The logarithmic convergence 
of $\rho_s$ for all $\alpha \le \alpha_c$ therefore in fact is the
expected stiffness scaling at a KT transition.
\end{references}
\end{document}